\documentclass[longauth]{aa}  

\usepackage{graphicx}
\usepackage{txfonts}
\usepackage{hyperref}
\usepackage{natbib}
\usepackage{booktabs}

\bibpunct{(}{)}{;}{a}{}{,}

\newcommand{\sophi}{{\sc SO/PHI}}
\newcommand{\hrt}{{\sc SO/PHI-HRT}}
\newcommand{\solo}{Solar Orbiter}

\newcommand{\kms}{${\rm km}\,{\rm s}^{-1}$}
\newcommand{\degree}{$^\circ$}
\makeatletter
\renewcommand*\aa@pageof{, page \thepage{} of \pageref*{LastPage}}

\usepackage[nolist,nohyperlinks]{acronym}
\begin{acronym}
\acro{MHD}{magnetohydrodynamic}
\acro{PSF}{point spread function}
\acro{RTE}{radiative transfer equations}
\acro{PD}{phase diversity}
\acro{CP}{circular polarization}
\acro{LP}{linear polarization}
\acro{FoV}{field of view}
\acro{RoI}{region of interest}
\acro{QS}{quiet Sun}
\end{acronym}

\begin{document} 

   \title{Spectropolarimetric investigation of magnetohydrodynamic wave modes in the photosphere: First results from PHI on board  Solar Orbiter}
   
   \author{D.~Calchetti\inst{1}\thanks{\hbox{Corresponding author: D.~Calchetti} \hbox{\email{calchetti@mps.mpg.de}}}\orcid{0000-0003-2755-5295}
     \and
   M.~Stangalini\inst{2}\orcid{0000-0002-5365-7546} \and 
   S.~Jafarzadeh\inst{1,3}\orcid{0000-0002-7711-5397} \and 
   G.~Valori\inst{1}\orcid{0000-0001-7809-0067}  \and 
   K.~Albert\inst{1}\orcid{0000-0002-3776-9548} \and 
   N. Albelo~Jorge\inst{1} \and 
   A.~Alvarez-Herrero\inst{4}\orcid{0000-0001-9228-3412} \and 
   T.~Appourchaux\inst{5}\orcid{0000-0002-1790-1951} \and 
   M.~Balaguer~Jiménez\inst{6}\orcid{0000-0003-4738-7727} \and 
   L.R.~Bellot~Rubio\inst{6} \orcid{0000-0001-8669-8857}  \and 
   J.~Blanco~Rodr\'\i guez\inst{7}\orcid{0000-0002-2055-441X} \and 
   A.~Feller\inst{1} \and 
   A.~Gandorfer\inst{1}\orcid{0000-0002-9972-9840} \and 
   D.~Germerott\inst{1} \and 
   L.~Gizon\inst{1,10}\orcid{0000-0001-7696-8665}\and  
   L.~Guerrero\inst{1} \and 
   P.~Gutierrez-Marques\inst{1}\orcid{0000-0003-2797-0392} \and 
   J.~Hirzberger\inst{1} \and 
   F. Kahil\inst{1}\orcid{0000-0002-4796-9527} \and 
   M.~Kolleck\inst{1} \and 
   A.~Korpi-Lagg\inst{1}\orcid{0000-0003-1459-7074} \and 
   A.~Moreno~Vacas\inst{6}\orcid{0000-0002-7336-0926} \and 
   D.~Orozco~Su\' arez\inst{6}\orcid{0000-0001-8829-1938} \and 
   I.~P\' erez-Grande\inst{8}\orcid{0000-0002-7145-2835} \and  
   E.~Sanchis Kilders\inst{7}\orcid{0000-0002-4208-3575} \and  
   J.~Schou\inst{1}\orcid{0000-0002-2391-6156} \and 
   U.~Sch\" uhle\inst{1}\orcid{0000-0001-6060-9078} \and 
   J.~Sinjan\inst{1}\orcid{0000-0002-5387-636X} \and 
   S.K.~Solanki\inst{1}\orcid{0000-0002-3418-8449}  \and 
   J.~Staub\inst{1}\orcid{0000-0001-9358-5834} \and 
   H.~Strecker\inst{6}\orcid{0000-0003-1483-4535} \and 
   J.C.~del~Toro~Iniesta\inst{6}\orcid{0000-0002-3387-026X}\and 
   R.~Volkmer\inst{9}\and 
   J.~Woch\inst{1}\orcid{0000-0001-5833-3738} 
   }

   \institute{
         Max-Planck-Institut f\"ur Sonnensystemforschung, Justus-von-Liebig-Weg 3, 37077 G\"ottingen, Germany \\ \email{solanki@mps.mpg.de}
         \and
         ASI, Italian Space Agency, Via del Politecnico snc, 00133, Rome, Italy
         \and
         Rosseland Centre for Solar Physics, University of Oslo, P.O. Box 1029 Blindern, NO-0315 Oslo, Norway
         \and
         Instituto Nacional de T\' ecnica Aeroespacial, Carretera de Ajalvir, km 4, E-28850 Torrej\' on de Ardoz, Spain
         \and
         Univ. Paris-Sud, Institut d’Astrophysique Spatiale, UMR 8617, CNRS, B\^ atiment 121, 91405 Orsay Cedex, France
         \and
         Instituto de Astrofísica de Andalucía (IAA-CSIC), Apartado de Correos 3004, E-18080 Granada, Spain \\ \email{jti@iaa.es}
         \and
         Universitat de Val\`encia, Catedr\'atico Jos\'e Beltr\'an 2, E-46980 Paterna-Valencia, Spain
         \and
         Instituto Universitario "Ignacio da Riva", Universidad Polit\'ecnica de Madrid, IDR/UPM, Plaza Cardenal Cisneros 3, E-28040 Madrid, Spain
         \and
         Leibniz-Institut für Sonnenphysik, Sch\" oneckstr. 6, D-79104 Freiburg, Germany
         \and
         Institut f\"ur Astrophysik, Georg-August-Universit\"at G\"ottingen, Friedrich-Hund-Platz 1, 37077 G\"ottingen, Germany
         }

\date{Received ***; accepted ***}

 
  \abstract
  {In November 2021, Solar Orbiter started its nominal mission phase. The remote-sensing instruments on board the spacecraft acquired scientific data during three observing windows surrounding the perihelion of the first orbit of this phase. }
  {The aim of the analysis is the detection of magnetohydrodynamic (MHD) wave modes in an active region by exploiting the capabilities of spectropolarimetric measurements. }
  {The High Resolution Telescope (HRT) of the Polarimetric and Helioseismic Imager (\sophi) on board the Solar Orbiter acquired a high-cadence data set of an active region. This is studied in the paper. B-$\omega$ and phase-difference analyses are applied on line-of-sight velocity and circular polarization maps and other averaged quantities.}
  {We find that several MHD modes at different frequencies are excited in all analysed structures. The leading sunspot shows a linear dependence of the phase lag on the angle between the magnetic field and the line of sight of the observer in its penumbra. The magnetic pore exhibits global resonances at several frequencies, which are also excited by different wave modes.}
  {The SO/PHI measurements clearly confirm the presence of magnetic and velocity oscillations that are compatible with one or more MHD wave modes in pores and a sunspot. Improvements in modelling are still necessary to interpret the relation between the fluctuations of different diagnostics.}

   \titlerunning{Spectropolarimetric investigation of \acs{MHD} wave modes with \sophi}
   \authorrunning{Calchetti D. et al.}
   \maketitle
%
%
\section{Introduction}\label{sec:intro}
\begin{figure*}[t]%
    \centering
    \includegraphics[width=\textwidth]{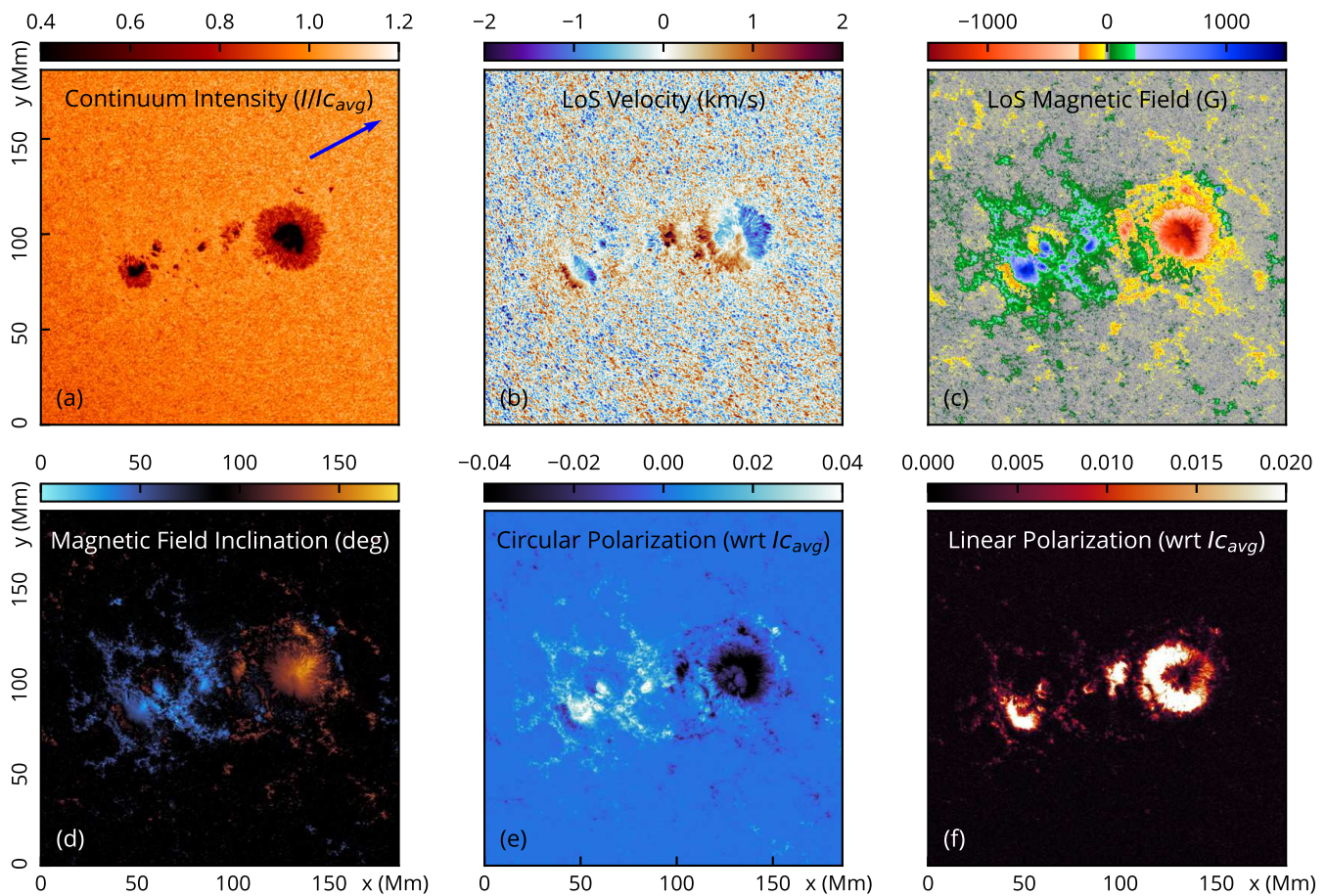}
    \caption{First image of the post-processed time series acquired on 7 March 2022 at 00:00:09UT. From panel \textit{(a)} to \textit{(f)}: Intensity in the continuum wavelength, LoS velocity, LoS magnetic field, inclination of the magnetic field vector, circular polarization, and linear polarization. Panels \textit{(b)} to \textit{(d)} are results of the RTE inversion with C-MILOS \citep{2007A&A...462.1137O}. Panels \textit{(e)} and \textit{(f)} show maps obtained with equation~\ref{eq:CP-LP}. The blue arrow in panel \textit{(a)} points towards the disc centre.}
    \label{fig:data}
\end{figure*}%
Wave phenomena in the solar atmosphere can traditionally be divided into regions dominated by five-minute oscillations (3~mHz, in the photosphere) and three-minute oscillations (5~mHz, in the chromosphere). 
This division is the result  of the density stratification and therefore of a cut-off frequency, thus making the solar atmosphere a filter for low-frequency waves below approximately 5~mHz  \citep[e.g.][]{2010ApJ...719..357F, 2015LRSP...12....6K}.

\cite{1968SoPh....4..286H} measured for the first time that the five-minute oscillations in umbrae and penumbrae of sunspots have a smaller amplitude than five-minute oscillations in the \ac{QS}. 
Many mechanisms have been suggested to explain this particular behaviour, such as the reduction in the efficiency of the excitation \citep{1977ApJ...212..243G,1988ApJ...326..462G} or the absorption of \textit{p}-mode oscillations \citep{1995ApJ...451..372C,2016ApJ...817...45R} and the subsequent conversion into \ac{MHD} modes. 
In addition to \textit{p}-mode conversion, recent studies have also revealed resonant oscillations in sunspots and pores. The resonant frequencies are dependent on the structure itself and on its shape \citep{2017ApJ...842...59J,2018ApJ...857...28K,2020NatAs...4..220J,b-omega,2022NatCo..13..479S,2022ApJ...938..143G}. 
These results have been obtained not only by analysing the intensity or velocity fluctuations, but in some cases, also by exploiting the potential of spectropolarimetric measurements. 
Spectropolarimetric data allow the inference of magnetic field properties \citep{2016LRSP...13....4D}, opening up new possibilities for the study of \ac{MHD} waves in the solar atmosphere. 
A peculiar case regarding this phenomenon was first reported by \cite{2011A&A...534A..65S}. 
They presented a magnetic pore that against expectations showed almost no fluctuations in the 3~mHz band in the photospheric layer, but showed clear oscillations at other frequencies. 
More recently, it was shown that these unexpected frequencies in the photosphere were resonant oscillations of the magnetic structure \citep{b-omega}. 

Spurious signals in polarimetric measurements usually originate from opacity effects or cross-talk between different Stokes parameters. 
Highly stable instruments are needed in order to disentangle these effects from real oscillations \citep{2000ApJ...534..989B}. 
One way to do this is by computing the phase-difference between different quantities, such as the line-of-sight (LoS) velocity field or the continuum intensity \citep{rueedi1998,fujimura2009,2018A&A...619A..63J}, and between different layers of the solar atmosphere \citep[e.g.][]{fujimura2009,2011ApJ...730L..37M,2021RSPTA.37900216S}. 
Only points with very high coherence should be considered in the analysis in order to guarantee a reliable relation between the signals. 
These phase values can also be compared to models of wave modes in magnetic flux tubes, such as those described in \cite{fujimura2009}, \cite{moreels2013}, and \cite{2013A&A...555A..75M,2015A&A...579A..73M}. 
Spectropolarimetric data are also necessary to constrain the excitation mechanisms of these wave modes \citep{2011SoPh..273...15V}. 

The solar magnetic field that is concentrated in flux tubes connects different layers of the solar atmosphere. 
Waves propagating in magnetic structures, or global resonant modes of the whole flux tube, can transport energy from the base of the solar atmosphere and dissipate it in the chromosphere or corona \citep{2015LRSP...12....6K,2017ApJS..229....7G,2017ApJS..229....9J,2017ApJS..229...10J,2018NatPh..14..480G,2020NatAs...4..220J,2020SSRv..216..140V}. 
\ac{MHD} waves can also cause changes in the plasma composition, such as the plasma fractionation \citep[e.g.][]{2015LRSP...12....2L,2021ApJ...907...16B,2021A&A...656A..87M}, thus enabling new ways to investigate the sources of the slow solar wind \citep[e.g.][]{2022ApJ...940...66B}. 

In March 2022, \solo\ \citep[SO][]{solo} started its first remote-sensing window of the nominal mission phase \citep{2020A&A...642A...3Z}. 
The Polarimetric and Helioseismic Imager \citep[PHI,][]{2020A&A...642A..11S} on board the \solo\ mission observed many different targets during the three remote-sensing windows of this orbit and provided photospheric spectropolarimetric data. 
\sophi\ returned full-Stokes vectors and/or inverted parameters at high resolution and high sensitivity. 
The stability of the instrument and the absence of the Earth's atmosphere play a key role in providing these high-quality data, which are also necessary for the detection of high-order and high-frequency \ac{MHD} wave modes \citep{2022NatCo..13..479S}. 

This paper further investigates the ability of detecting real magnetic field oscillations \citep[e.g.][]{2021RSPTA.37900216S,2021RSPTA.37900182K} associated with particular \ac{MHD} wave modes  (e.g. \citealt{roberts2019}; see \citealt{2023LRSP...00..000J} for a recent review). 
We describe the first high-cadence observations taken by \sophi. 
A Fourier analysis is performed on a magnetic field diagnostic and the LoS velocity to detect \ac{MHD} wave modes in the sunspots and pores of the observed active region. 
This investigation has been conducted by studying the B-$\omega$ diagrams of and the phase lags between these quantities. 
The results show magnetic and velocity oscillations that are compatible with one or more \ac{MHD} wave modes. 
%
%
\begin{figure*}[t]
    \centering
    \includegraphics[width=\textwidth]{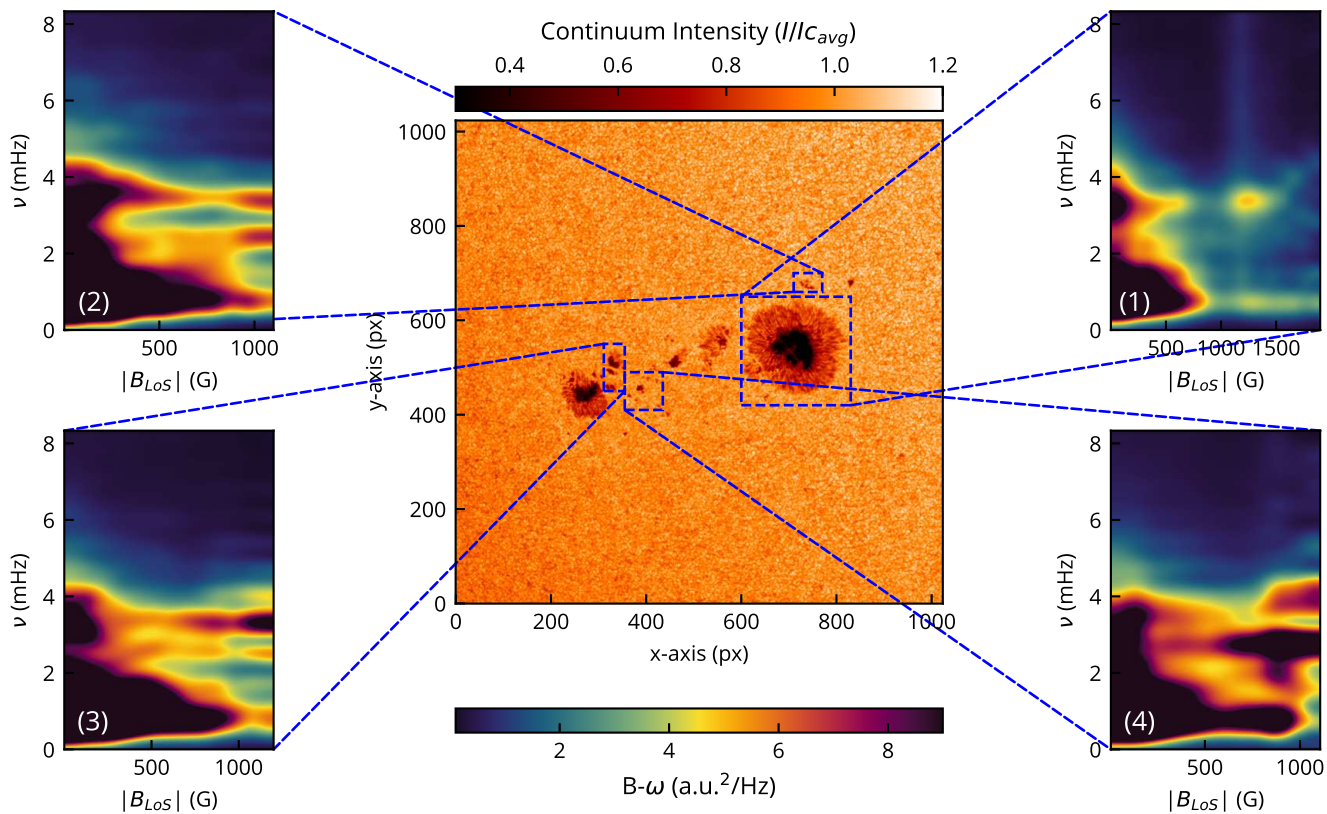}
    \caption{Results of the B-$\omega$ analysis of the LoS velocity field. \textit{Central panel}: Continuum intensity map showing the four RoIs. \textit{Panels (1-4)}: B-$\omega$ diagrams of the PSD of the LoS velocity field in the corresponding RoIs. The absolute value of the LoS magnetic field is plotted on the $x$ -axes, and the frequency is plotted on the $y$ axes. The same colour scale has been used for all these panels and is shown below the central panel. We have chosen the saturation levels in order to emphasise the power increases in the magnetic regions. }
    \label{fig:b-w-vlos}
\end{figure*}
\section{Observations}\label{sec:obs}%
\begin{figure*}[t]
    \centering
    \includegraphics[width=\textwidth]{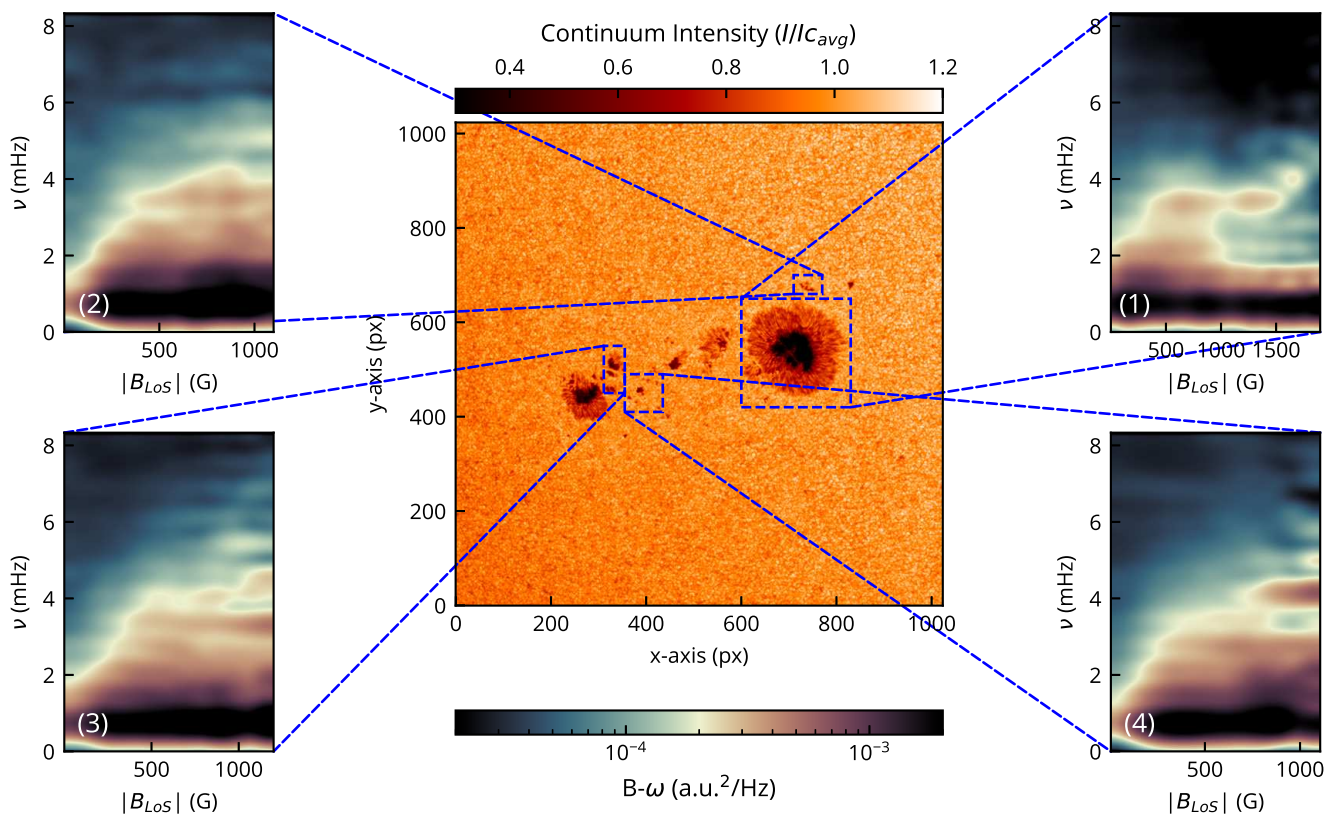}
    \caption{Results of the B-$\omega$ analysis on the CP time series. \textit{Central panel}: Same as for Fig.~\ref{fig:b-w-vlos}. \textit{Panels (1-4)}: B-$\omega$ diagrams of the PSD of the \acs{CP} time series in the corresponding RoIs. The absolute value of the LoS magnetic field is plotted on the $x$ -axes, and the frequency is plotted on the $y$ -axes. The same colour scale has been used for all these panels and is shown below the central panel. We have chosen the saturation levels in order to emphasise the power increases in the magnetic regions. $a.u.$ indicates arbitrary units.}
    \label{fig:b-w-CP}
\end{figure*}
\subsection{Observational data}\label{subsec:data}
The data set analysed in this work was acquired by the \sophi\ High Resolution Telescope \citep[HRT,][]{gandorfer2018high} for the Nanoflares Solar Orbiter Observing Plan\footnote{more information can be found here: \url{https://issues.cosmos.esa.int/solarorbiterwiki/display/SOSP/SOOP+pages}} during the inferior conjunction on 7 March 2022 from 00:00:09 to 00:45:09 UT. The spacecraft position was exactly in between the Earth and the Sun at that time. 

This time series consists of a 45-minute observation with a cadence of 60 seconds, sampling the active region (AR) with NOAA number 12960. 
The distance from the Sun at the time of observation was 0.501~au, resulting in a pixel scale of 181 km at disc centre on the \hrt\ detector. 
The AR was observed at a cosine of heliocentric angle ($\mu$) of 0.87, located at $-31.2$\degree\ in longitude and $-17.5$\degree\ in latitude in heliographic Stonyhurst coordinates (see \citealt{2006A&A...449..791T} for more details about solar coordinates). 
\hrt\ acquired 45 spectral-line scans at six wavelength positions (five points within the Fe~{\sc i} line centred at 617.3~nm, and one point in the near continuum), all with four different polarimetric modulation states. 
The exposure time of each individual frame is 6~ms, and individual images result from the accumulation over 16 exposures to reach the required S/N. The total cadence for obtaining a single data set of 24 images is 60 seconds. 
\subsection{Data reduction}\label{subsec:red}
The data were down-linked as raw and were then processed and calibrated using the \hrt\ on-ground pipeline \citep{hrt-pipeline}. The full-Stokes vector at each wavelength and the results of the inversion of the \ac{RTE} in Milne-Eddington approximation were obtained with C-MILOS \citep{2007A&A...462.1137O}. 

A new processing step was added to the data calibration pipeline prior to the \ac{RTE} inversion for this work compared to \cite{hrt-pipeline}. The step is currently being included into the standard version of the pipeline.
We retrieved the \ac{PSF} of the instrument with a \ac{PD} analysis (\citealt{fatima-PD}, based on \citealt{1992JOSAA...9.1072P,1994A&AS..107..243L}). 

The images were corrected for optical aberrations introduced mostly by the entrance window of the instrument. 
For this restoration, called aberration correction, the images are deconvolved by a reduced \ac{PSF}. 
The latter is obtained by dividing the full \ac{PSF} obtained from the \ac{PD} calibration by the pure diffraction-limited \ac{PSF} of a circular aperture (Airy function) corresponding to the ideal (aberration-free) entrance pupil of the instrument. 
This convolution is done mainly to reduce the additional noise introduced by the deconvolution algorithm. 
More information about the \ac{PD} analysis and retrieval of the wavefront error of the \hrt\ can be found in \cite{fatima-PD,fatima-PD2}. 

The noise level obtained at the end of the processing is $1.8\times10^{-3}$ in Stokes V and Q and $2.2\times10^{-3}$ in Stokes U in units of the continuum intensity ($Ic_{avg}$). 
For operational reasons, the image stabilisation system \citep[ISS,][]{ISS} of the instrument was off during the acquisition. 
One of the main functions of the ISS is to prevent motion-induced polarimetric cross-talk and improve polarimetric precision that can be affected by the spacecraft jitter. 
However, they can only be partially compensated for by the data reduction pipeline. 
Therefore, the absence of the ISS and the deconvolution algorithm cause the higher noise with respect to the nominal level of $10^{-3}$ \citep[see][]{hrt-pipeline,jonas-2023}. 

The spacecraft jitter induces movements of the image on the detector plane. 
The solar scene is therefore sampled at different spectral points due to the cavity errors of the etalon, which are anchored to the detector plane. 
The effect of this shift in the sampling wavelength during the accumulations of each frame was tested using synthetic Stokes profiles of the Fe~{\sc i} line with non-local thermodynamic equilibrium (NLTE) effects \citep[described in][]{Smitha2023}. 
The effects are negligible and do not produce any artefacts in the data. 

In addition to the processing and calibration procedures described above, we decided to post-process the full Stokes vectors in order to obtain a time series ready for scientific analysis. 
First, we selected a sub-region of $1024\,\times\,1024$~px, which includes the whole AR. 
We then co-aligned the image time-series using a cross-correlation algorithm with sub-pixel accuracy \citep[based on][]{2008OptL...33..156G} that is applied to the continuum intensity maps. 
Once we modulated the Stokes vectors using the inverted demodulation matrix, we performed a pixel-by-pixel linear interpolation between consecutive data sets to produce pseudo co-temporal modulated data in which all the polarimetric and wavelength scans, that contribute to a given set of Stokes profiles, have the same observation time \citep[following the technique by][]{2017ApJ...849....7O}.
Lastly, we demodulated and corrected for cross-talk to remove any possible spurious effects from the interpolation of modulated polarimetric frames and then we ran the \ac{RTE} inversion of the pseudo co-temporal full-Stokes vectors with C-MILOS. 
The fully reduced full-Stokes vectors were also used to compute \ac{CP} and \ac{LP} maps following the definitions by \cite{2011SoPh..268...57M},
\begin{equation}\label{eq:CP-LP}
CP = \frac{1}{4} \sum_{i=1}^4 a_i V_i\,\,\text{and}\,\,LP = \frac{1}{4} \sum_{i=1}^4 \sqrt{Q_i^2 + U_i^2}\,,
\end{equation}
where $Q_i$, $U_i$, and $V_i$ are the polarized components of the Stokes vectors in the blue ($i=[1,2]$) and red ($i=[3,4]$) line wings, and $a = [1,1,-1,-1]$. 
One example of the data products for the first time step of the time series is shown in Fig.~\ref{fig:data}. 
The full \ac{FoV} shown in all the panels is $1024\,\times\,1024$~px, and the maps in panels \textit{(b-d)} are results of the \ac{RTE} inversion. 
The continuum intensity and the \ac{CP} and \ac{LP} maps in panels \textit{(a), (e)} and \textit{(f)} are shown in units of the continuum intensity. 
Negative velocities in panel \textit{(b)} are flows towards the observer, and angles smaller than 90\degree\ in panel \textit{(d)} point towards the observer.  
%
%
\section{Methods and desults}
We investigated the velocity and magnetic oscillations in AR12960. 
We decided to focus on the LoS velocity and on a magnetic diagnostic, such as \ac{CP} instead of the LoS magnetic field. 
We made this decision to avoid the higher spatial noise from the \ac{RTE} inversion and to have a decreased noise through the spectral average that was performed to obtain the \ac{CP} maps. 
A similar choice was also made by  \cite{2011ApJ...730L..37M,2013A&A...549A.116J,b-omega,2021NatAs...5..691S}, for instance. 
A Fourier analysis, pixel by pixel, was carried out to study the oscillations in the whole \ac{FoV}. 
Furthermore, we obtained the power spectral density (PSD) in each pixel of the time series with a highest detectable frequency of 8.33~mHz. 

In order to investigate the dependence of the velocity and magnetic dynamics on the photospheric magnetic field, we constructed B-$\omega$ diagrams \citep{b-omega} of different magnetic structures in the \ac{FoV}. 
This technique is particularly suited to find coherent oscillations in magnetic structures with irregular shapes because it is not directly dependent on the spatial extension of the magnetic field, unlike the k-$\omega$ analysis \citep[e.g.][]{1983Natur.302...24D} or the power-distance diagram \citep{2013A&A...560A..84S}. 
The B-$\omega$ diagram was obtained by dividing each \ac{RoI} into magnetic bins (100~G each in this case) and by averaging the PSDs of the chosen quantity (LoS velocity and \ac{CP} in this work) in each magnetic bin. 
To do this, we used the absolute values of the LoS magnetic fields for the binning. 
In this case, the LoS magnetic field was preferred to \ac{CP} as a more straightforward quantity when binning the chosen \ac{RoI}. 

The B-$\omega$ diagrams of the LoS velocity field and of the \ac{CP} time series are illustrated in Figs.~\ref{fig:b-w-vlos} and \ref{fig:b-w-CP}, respectively. 
We present four of the magnetic structures in the \ac{FoV} in this paper. 
They show the most interesting wave modes. 
\ac{RoI} (1) represents the leading spot of AR12960. 
Its velocity field clearly shows 3~mHz oscillations in the \ac{QS} region, and  two small high-power regions are evident for magnetic field values above 1000~G. 
Nevertheless, smaller peaks are present in the umbra of the sunspot at 3~mHz and 4~mHz for magnetic field values higher than 1500~G. 
On the other hand, the B-$\omega$ diagram of \ac{CP} of the same region in Fig.~\ref{fig:b-w-CP} shows similar peaks, but the overall distribution of the power is different, for example the small increase in power at 4~mHz in the umbra. 
The other three regions represent small pores next to the leading and trailing sunspots. 
In all of these cases, different oscillation modes in the \ac{QS} and the magnetic regions are clearly visible, which shows very peculiar modes in different magnetic field ranges. 
Differences are also visible between the LoS velocity and \ac{CP} diagrams. 
In particular, the ratio of the amplitudes of the power peaks are different for each wave mode that is detected in the magnetic structures. 
\begin{figure}[t]
    \centering
    \includegraphics[width=\columnwidth]{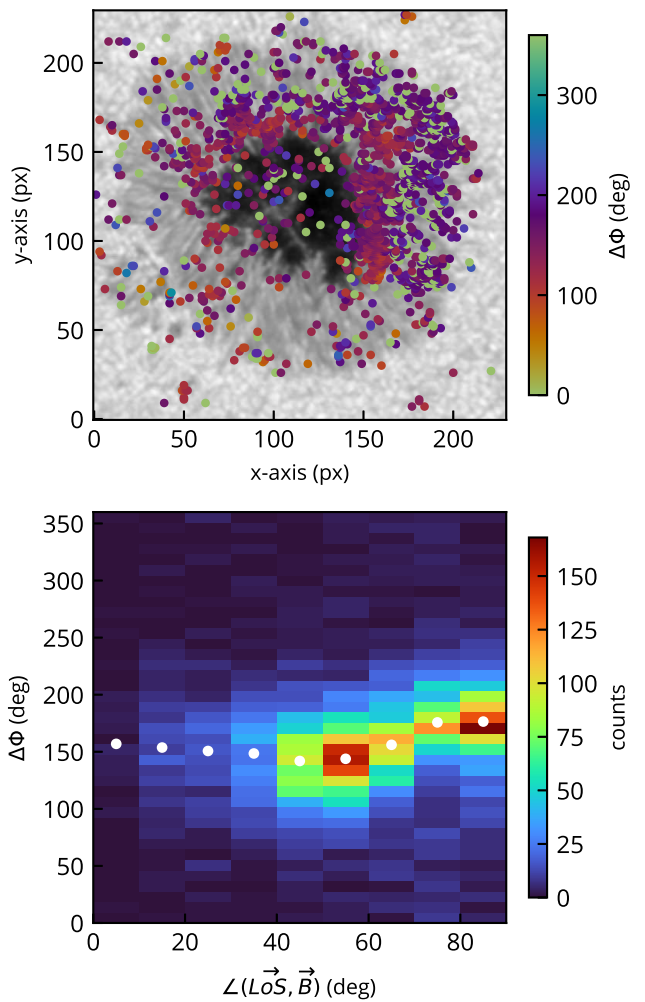}
    \caption{Results of the Fourier analysis for \ac{RoI} (1). \textit{Top panel}: Continuum intensity of \ac{RoI} (1). The phase-differences for the pixels show a high coherence between the LoS velocity and \ac{CP} time series at 3.3 mHz that is over-plotted. The colour scale represents the values of the phase difference in each considered point. \textit{Bottom panel}: 2D histogram as a function of the angle between the magnetic field vector and the LoS (x-axis) and the phase difference between the LoS velocity and \ac{CP} time series at 3.3~mHz. The same points as in the upper panel have been used to build this diagram. The withe dots represent the central position obtained with a Gaussian fit of the phase-difference value at different angle bins.}
    \label{fig:vlos-cp-roi1}
\end{figure}
In order to further investigate the oscillation modes shown in the B-$\omega$ diagrams, we performed two different analyses depending on the type of structure. 
In \ac{RoI} (1), we carried out a pixel by pixel phase-difference analysis between the LoS velocity and the \ac{CP} series within the leading sunspot. 
In \acp{RoI} (2), (3), and (4), we computed the average signals of the LoS velocity, \ac{CP}, continuum intensity, and cross section of the magnetic structures and we used those to detect \ac{MHD} modes. 
This choice was motivated mostly by the size of the analysed structures. 
First, a pixel-by-pixel analysis is better suited for larger regions, such as the leading sunspot. 
Second, the averaged analysis is less well suited for the sunspot due to the definition of the cross section, which includes the umbra and not the penumbra. 

Fig.~\ref{fig:vlos-cp-roi1} shows results obtained for the sunspot in \ac{RoI} (1). 
The upper plot illustrates some of the phase differences between the LoS velocity and the \ac{CP} signal obtained from a Fourier analysis, over-plotted on a continuum intensity map. 
A frequency range centred on 3.3 mHz (with 1 mHz width) was considered in order to include most of the oscillating power detected with the B-$\omega$ diagram. 
This analysis was performed pixel by pixel in locations in which the continuum intensity was lower than 0.9 in order to exclude \ac{QS} and plage regions. 
The points represented here are only those with an average coherence over the frequency bin higher than 90\%. 
The coherence between the signals represents a linear relation between them and it is not directly related to the amplitudes or the phases of the signals. 
The high coherence ensures high reliability in the phase measurements between the two series \citep{2017ApJ...835..148V,2018ApJ...869..110S}. 

The lower panel of Fig. \ref{fig:vlos-cp-roi1} shows a 2D histogram of the phase-difference obtained from the analysis. 
The background colour displays the histogram density (i.e. counts), plotting the phase difference between the LoS velocity and the \ac{CP} signal as a function of the angle between the vector magnetic field and the LoS. 
This analysis was made by retrieving the vector magnetic field from the C-MILOS inversion and by disambiguation of the azimuth of the magnetic field with the minimal energy method \citep{metcalf1994} adapted from the SDO/HMI pipeline \citep{2014SoPh..289.3483H}. 
The angle between the LoS and the magnetic field was obtained using equation 1 in \cite{1990SoPh..126...21G}. 
We also performed a Gaussian fit of the phase-lag distribution at each angle bin to obtain the central phase difference at that specific interval. 
The mean values of the fit are plotted as white dots in the same diagram. 
The diagram clearly shows a change in the phase difference from 140\degree\ to 180\degree\ in the angle range between 40\degree\ and 90\degree. 

\begin{figure}[t]
    \centering
    \includegraphics[width=0.8\columnwidth]{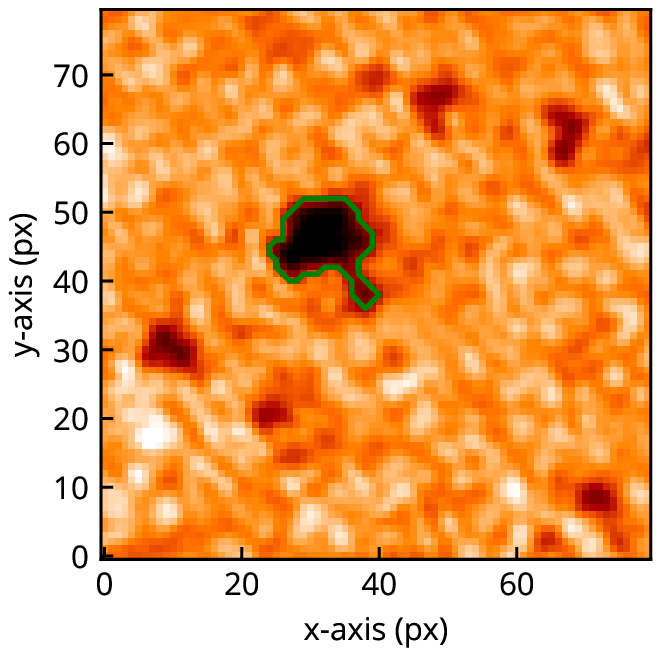}
    \includegraphics[width=\columnwidth]{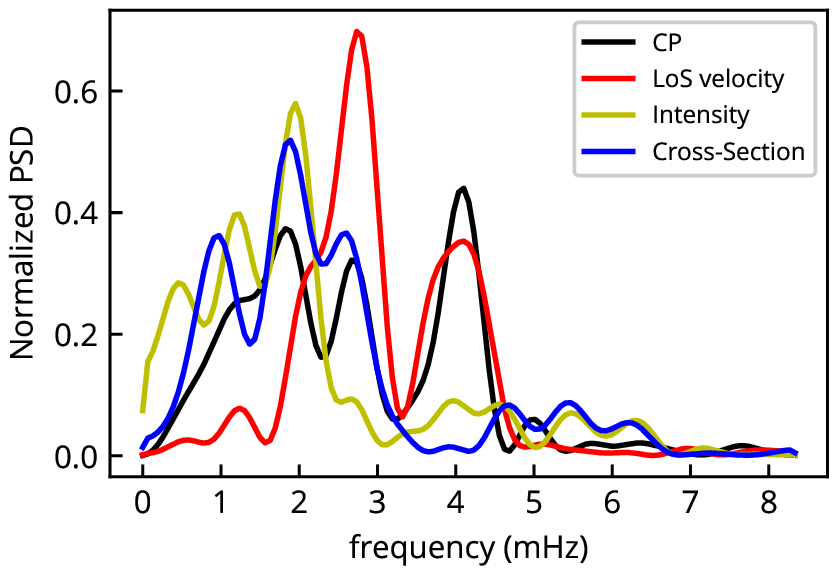}
    \caption{Results of the Fourier analysis of \ac{RoI} (4). \textit{Top panel}: Continuum intensity of \ac{RoI} (4). The green contour shows the tracked region at this specific time step. \textit{Bottom panel}: PSDs obtained by averaging the signal in the pore in \ac{RoI}~(4). \textit{Black line}: \acs{CP}. \textit{Red line}: LoS velocity field. \textit{Yellow line}: Continuum intensity. \textit{Blue line}: Cross section. The PSDs have been normalized to their total power (i.e. frequency integrated).}
    \label{fig:roi4}
\end{figure}

A different approach was pursued on smaller magnetic structures in {RoIs} (2), (3), and (4). 
We tracked each magnetic structure as in \cite{morton2011}. 
We took the median value of the continuum intensity in a \ac{QS} region next to the \ac{RoI} and considered only pixels with values at least 3$\sigma$ below the median. 
This technique provides a 99\% confidence level that the dark magnetic pores or umbrae are contoured. 
The most interesting result is obtained from \ac{RoI} (4), which is represented in Fig.~\ref{fig:roi4}. 
The same plots for \acp{RoI} (2) and (3) are shown in Fig. \ref{fig:roi2} and \ref{fig:roi3}. 
The bottom panel of Fig.~\ref{fig:roi4} shows the PSDs of four different quantities from averaged signals in the contoured magnetic pore at each time step (i.e. \ac{CP}, LoS velocity, continuum intensity, and the cross section of the pore). 
The cross section is defined as the surface inside the bounds of the contour. 
The PSDs were normalized by dividing them by their total power. 
This was done mostly for representation purposes. 
This plot clearly shows three main peaks that are common to all the PSDs at 1.9, 2.9, and 4.1~mHz. 
In Table \ref{tab:ph} we show the phase differences between these quantities, computed for the frequencies corresponding to the maxima of the PSDs. 
The phase differences at 1.9~mHz and 2.9~mHz show similar values, but this is different for the other peak at 4.1~mHz. 
This last peak is dominated by oscillations in \ac{CP} and LoS velocities, whereas the continuum intensity and cross section have almost negligible amplitude. 
\begin{table}[t!]
    \caption{Phase differences between averaged \acs{CP}, LoS velocity (v$_{LoS}$), cross section (CS), and continuum intensity (Icnt) at 1.9, 2.9, and 4.1 mHz in the magnetic pore in \ac{RoI} (4). The last three values at 4.1~mHz are not shown because CS and Icnt have low signal at this frequency. }
    \centering
    \begin{tabular}{lrrr}
        \toprule
         & 1.9 mHz & 2.9 mHz & 4.1 mHz \\
        \midrule
        CP - v$_{LoS}$ & -62\degree & -65\degree & -81\degree \\
        \midrule
        CS - Icnt & 28\degree & 24\degree & * \\
        \midrule
        Icnt - CP & 6\degree & -22\degree & * \\
        \midrule
        v$_{LoS}$ - Icnt & 65\degree & 83\degree & * \\
        \bottomrule
    \end{tabular}
    \label{tab:ph}
\end{table}%
%
%
%
\section{Discussion}
The analysis performed on this first high-cadence time series acquired by \hrt\ has shown various \ac{MHD} modes in different magnetic structures in AR12960. 
The B-$\omega$ diagrams in Fig.~\ref{fig:b-w-vlos} and \ref{fig:b-w-CP} reveal a separation between the \ac{QS} and the magnetic structures. 
While the power of the LoS velocity in the \ac{QS} always peaks around 3.3 mHz, the oscillations in regions with LoS magnetic fields higher than 500 G show several different patterns. 
The diagrams in Fig.~\ref{fig:b-w-CP} also show different frequencies of oscillations in each structure, with only low frequency power in the \ac{QS} regime, as in Fig. \ref{fig:b-w-vlos}: This is likely generated by the noise in the data. 
However, we cannot exclude a physical source such as plasma motions, granulation, or even internal gravity waves, which can be distinguished by their different properties in a phase-difference diagram \citep[e.g.][]{2017ApJ...835..148V,2021RSPTA.37900178C}. 
The B-$\omega$ analysis remains a powerful tool for investigating the presence of wave modes in these structures, allowing the detection of a clear separation between \ac{QS} oscillation patterns and the global eigenmodes of the magnetic flux tubes in each \ac{RoI}. 

The phase relation between the LoS velocity and \ac{CP} was taken into consideration in the case of the leading sunspot in \ac{RoI} (1). 
The relation of the phase difference to the angle between the magnetic field vector and the LoS of the observer is particularly interesting, as shown in Fig.~\ref{fig:vlos-cp-roi1}. 
This analysis is motivated by an increase in power and coherence in regions of the penumbra that point towards the disc centre. 
This difference in the inclination angle of the magnetic field vector is also visible in the increased amplitude of the \ac{CP} and is reduced in \ac{LP} in Fig. \ref{fig:data}. 
The behaviour of magneto-acoustic waves in an environment with an inclined magnetic field has broadly been investigated by many authors \citep[e.g.][]{2006MNRAS.372..551S}. 
The mode transmission and conversion strongly depend on the attack angle of the wave, which is the angle between the wave vector and the magnetic field vector. 
As an effect of the polarization of the slow wave modes \citep[which are expected to have a small attack angle; see e.g.][]{2015SSRv..190..103J}, we would expect to observe this type of waves along inclined magnetic field lines all around the penumbra of the sunspot. 
However, Fig.~\ref{fig:vlos-cp-roi1} shows a linear relation between the phase difference between the LoS velocity and \ac{CP} and the angle between the LoS of the observer and the magnetic field vector. 
This relation is evident in the phase-lag window between 140\degree\ and 180\degree\ and in the separation angle between 40\degree\ and 90\degree. 
The top panel in Fig.~\ref{fig:vlos-cp-roi1} clearly shows that most of the points with high coherence lie in the region of the penumbra towards the disc centre, and most of the points with smaller phase lag are close to the umbra-penumbra boundary. 
We would also expect high coherence between the \ac{CP} and LoS velocity in a slow magneto-acoustic mode, but there is no information regarding their phase relation in the literature. 

One possible explanation for this variation in the phase difference might be the variation in the height difference between the \ac{CP} and the LoS velocity maps. 
When we consider the same height in the \ac{QS} with a phase difference of 180\degree, we would expect a height difference at the umbra-penumbra boundary of \begin{equation}\Delta h = \frac{c_s\Delta\Phi}{2\pi\nu},\end{equation} where $c_s$ is the sound speed, $\nu$ is the frequency, and $\Delta\Phi$ is the phase difference. 
Considering $c_s = 7$~\kms, $\nu = 3$~mHz, and $\Delta\Phi = 40$\degree \ (which is the change in the phase difference as reported in the lower panel of Fig. \ref{fig:vlos-cp-roi1}), we obtain $\Delta h \approx 250$~km. 
According to \cite{2011SoPh..271...27F}, the formation height of the line core of the Fe~{\sc i} line is $\sim$250 km, so that we would not expect a height difference comparable to this value. 
This is also due to the response functions to the LoS velocity in C-MILOS presented by \cite{2007A&A...462.1137O}, which are more peaked in the wings of the line. 
One more possible interpretation of this result is that the propagation speed of the waves changes because of the different physical properties of the photospheric plasma in the sunspot penumbra. 
This might explain a variation in the phase difference, but further investigations are needed to understand this behaviour better. 

For the small pores and spots in \acp{RoI} (2), (3), and (4), we performed an analysis similar to \cite{morton2011} to detect \ac{MHD} modes in these magnetic flux tubes. 
The bottom panel of Fig. \ref{fig:roi4} shows the PSDs obtained on averaged quantities in a small magnetic pore. 
The results obtained from the phase analysis (Table \ref{tab:ph}) can be used to infer which type of wave mode dominates at each frequency \citep[see][]{moreels2013,2013A&A...555A..75M}. 
The two peaks at 1.9~mHz and 2.9~mHz show very similar phase differences, particularly between the \ac{CP} and LoS velocity and between the cross section and continuum intensity. 
According to the findings of \cite{moreels2013}, these values can represent a slow standing surface mode, and the peaks might be harmonics of the same mode, even though the averaged PSDs show different power ratios at different peaks. 
It is also important to consider that the observed phase lags are not exactly identical to those predicted by \cite{moreels2013}. 
This is expected, because their values were found for a straight cylinder with constant plasma parameters inside and outside the flux tube. 
Moreover, only one wave mode was excited in the flux tube simulated by \cite{moreels2013}, whereas at least two different modes occur here at the same time in the same magnetic pore. 
The wave mode detected at 4.1 mHz shows different phase lags and a very low oscillatory power in the cross section and continuum intensity. 
This indicates a different wave mode in the same magnetic structure that then resonates at different frequencies and with different \ac{MHD} modes at the same time. 
According to the simplified model described in \cite{fujimura2009}, the phase lag between \ac{CP} and LoS velocity and the small oscillating power in intensity and cross section might have been generated by a non-compressible kink mode. 
In particular, this might be caused by a mix of ascending and descending propagating modes or by a standing kink mode. 

Furthermore, it is worth noting that the frequency peaks found in the spatially averaged LoS velocity and \ac{CP} time series are the same as we found in the B-$\omega$ diagrams in Fig. \ref{fig:b-w-vlos} and \ref{fig:b-w-CP}. 
The same result is obtained by averaging the signals and the PSDs in the same magnetic structures. 
This shows that a coherent and global oscillation of the flux tube generated by a given wave mode has been detected with two different techniques. 
%
%
\section{Summary and conclusion}
We reported the detection of \ac{MHD} modes in AR12960 using spectropolarimetric data acquired by \hrt\ during the first remote-sensing window of \solo\ on 7 March 2022 during the inferior conjunction. 
The 45-minute time series was calibrated with the \hrt\ on-ground pipeline and was then post-processed with the procedures described in Section~\ref{subsec:red}. 
Magnetohydrodynamic modes were detected in the leading sunspot of the AR and in several small magnetic structures. 
The B-$\omega$ diagrams of the \acp{RoI}, shown in Fig.~\ref{fig:b-w-vlos} and \ref{fig:b-w-CP}, were used to detect the wave modes at different frequencies and magnetic field ranges. 
The leading sunspot in \ac{RoI} (1) and the pore in \ac{RoI} (4) show a clear relation in the oscillations between different quantities and diagnostics. 

We detected a high coherence between oscillations in the \ac{CP} and LoS velocity in the sunspot in \ac{RoI} (1), with a phase lag ranging from 140\degree\ to 180\degree. 
These lags depend approximately linearly on the angle of separation between the magnetic field vector and the LoS of the observer. 
The reason for this dependence is not fully understood yet, but we can exclude the height difference between the two time series as the source of the effect. 

The small pore in \ac{RoI} (4) was tracked with the technique presented in \cite{morton2011} to measure the oscillation of the whole structure in the LoS velocity, \ac{CP}, intensity, and cross section. 
The identified modes in which the oscillation power peaked at 1.9~mHz and 2.9~mHz are compatible with slow standing sausage modes according to \cite{moreels2013}, whereas the peak at 4.1 mHz might be generated by a non-compressible kink mode. 
These modes are compatible with the frequencies observed in the B-$\omega$ diagrams and demonstrate the possible presence of multiple \ac{MHD} modes in magnetic structures in the solar photosphere. 

In the future, a longer time series with higher spatial resolution and signal-to-noise ratio would facilitate the detection of wave modes at higher frequencies and higher-order modes. 
Co-observations from different vantage points with ground-based and Earth-orbiting instruments can also be extremely important to measure the ratio of the horizontal to vertical displacement in \textit{p-} and \textit{f-} modes \citep[e.g.][]{jesper2023} or to distinguish longitudinal and transverse oscillations. 
%
%

\begin{acknowledgements}
D.~C. would like to thank H.~N.~Smitha for her valuable inputs. 
Solar Orbiter is a space mission of international collaboration between ESA and NASA, operated by ESA.  We are grateful to the ESA SOC and MOC teams for their support. 
The German contribution to SO/PHI is funded by the BMWi through DLR and by MPG central funds. 
The Spanish contribution is funded by AEI/MCIN/10.13039/501100011033/ (RTI2018-096886-C5, PID2021-125325OB-C5, PCI2022-135009-2) and ERDF “A way of making Europe”; “Center of Excellence Severo Ochoa” awards to IAA-CSIC (SEV-2017-0709, CEX2021-001131-S); and a Ramón y Cajal fellowship awarded to DOS. 
The French contribution is funded by CNES. 
The authors wish to acknowledge scientific discussions with the Waves in the Lower Solar Atmosphere (WaLSA; \href{https://WaLSA.team}{www.WaLSA.team}) team, which has been supported by the Research Council of Norway (project no. 262622), The Royal Society (award no. Hooke18b/SCTM), and the International Space Science Institute (ISSI Team 502).
\end{acknowledgements}
%
%
\bibliographystyle{aa}
\bibliography{bibfile}
%
%
\begin{appendix}
\section{Results from RoIs (2) and (3)}
\begin{figure}[h]
    \centering
    \includegraphics{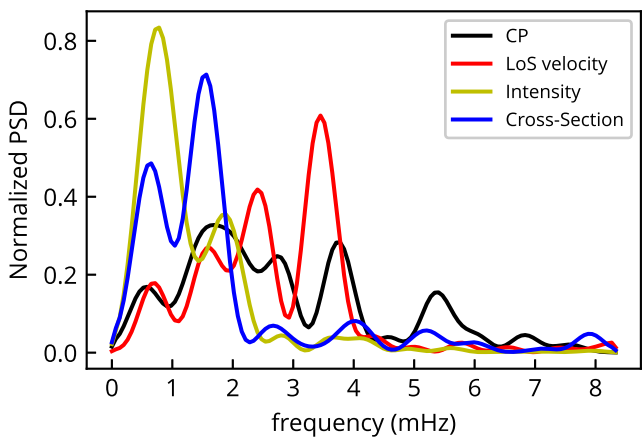}
    \caption{PSDs obtained by averaging the signal in the pore in \ac{RoI}~(2). \textit{Black line}: \acs{CP}. \textit{Red line}: LoS velocity field. \textit{Yellow line}: continuum intensity. \textit{Blue line}: Cross section. The PSDs have been normalized to their total power.}
    \label{fig:roi2}
\end{figure}

\begin{figure}
    \centering
    \includegraphics{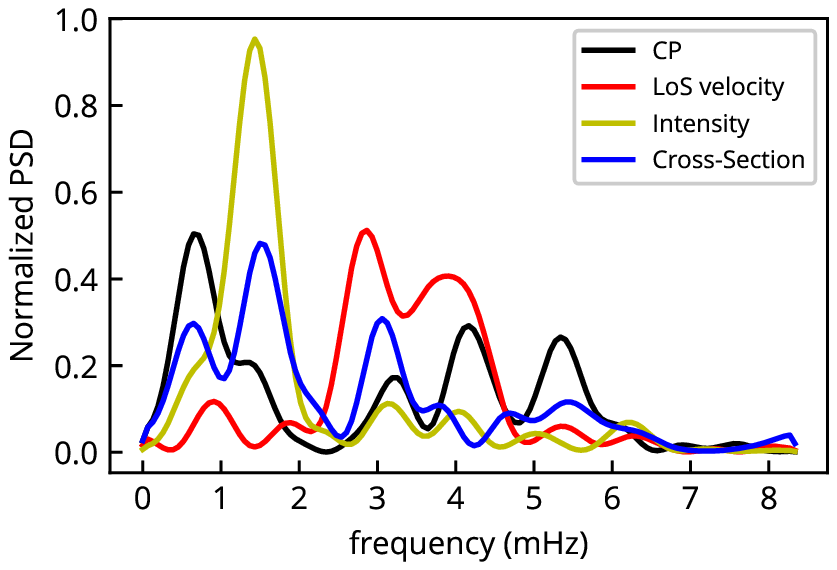}
    \caption{PSDs obtained by averaging the signal in the pore in \ac{RoI}~(3). \textit{Black line}: \acs{CP}. \textit{Red line}: LoS velocity field. \textit{Yellow line}: continuum intensity. \textit{Blue line}: Cross section. The PSDs have been normalized to their total power.}
    \label{fig:roi3}
\end{figure}

The analysis performed on \ac{RoI} (4) (shown in Fig. \ref{fig:roi4}) was also carried out on \acp{RoI} (2) and (3) (see Fig. \ref{fig:b-w-vlos} and \ref{fig:b-w-CP} for the context and B-$\omega$ diagrams). 
The PSDs of the averaged quantities obtained after the tracking of the magnetic structures is shown in Fig. \ref{fig:roi2} and \ref{fig:roi3} for \acp{RoI} (2) and (3), respectively. 

The PSDs show several peaks, particularly in the \ac{CP} and LoS velocity, whereas the continuum intensity and the cross section are dominated by low-frequency oscillations. 
In the same way as for \ac{RoI} (4), but much more pronounced, the power peaks between different quantities shift weakly. 
For this reason, we did not perform a phase analysis for any of the time series for \acp{RoI}~(2) and (3). 
This type of behaviour was also reported in \cite{b-omega}. 
The authors suggested a mix of slow and fast wave modes or different formation heights of the signals as a possible explanation, but a clear answer is still lacking. 
\end{appendix}
\end{document}